# Tie-line Security Regions in High Dimension for Renewable Accommodations

Wei Lin, *Member, IEEE*, Hua Jiang, Zhifang Yang, *Member, IEEE*

*Abstract*—Tie-line power exchanges among regional power systems facilitate renewable accommodations. Power exchanges can be calculated via a tie-line security region that provides the feasible region of the coupling parameters among regional power systems. However, a tie-line security region is a high-dimension polytope due to multiple time periods and border buses inherently in power system operations, leading to the considerable computational burden. A fast calculation method for tie-line security regions in high dimension is studied in this paper. The high-dimension polytope across all the time periods is decomposed as a Cartesian production of lower-dimension polytopes at each time period by leveraging dispatch levels of generations. For each lower-dimension polytope, the computational burden brought by multiple border buses is alleviated by aggregating tie-line power. Also, minimum renewable curtailments are preserved by incorporating an additional dimension in the tie-line security region. For the coupling parameters located within our tie-line security region, a feasible decision of the regional power system exists. Finally, the tie-line security region is used to reduce renewable curtailments in an interconnected power system under a decentralized and non-iterative framework. The performance of the presented methods is corroborated in the IEEE 9-bus system, a 661-bus utility system and a five-region system.

## I. Introduction

The decarbonization of power systems has drawn much attention to resolve the environmental issues brought by the excessive use of fossil fuels [1]. Consequently, the installed capacities of renewables have been fast increasing around the world [2]. Within this content, renewable accommodations via tie-line power exchanges have been important in interconnected power systems, as observed around the world [3]. For example, the high-penetration renewables in Europe are accommodated via tie-line power exchanges among different countries (e.g., Norway, Denmark) [4]. The hydroelectricity in Canada is accommodated via North American interconnections [5].

Generally, different regional power systems are operated by different operators. Consequently, the tie-line power exchanges to facilitate renewable accommodations would be difficult to be determined under a centralized framework. As an alternative, the tie-line power exchanges can be determined via the power transfer capability that has been required in China, Europe, and America [6]. Essentially, the power transfer capability can be described as a tie-line security region in the domain of the coupling parameters among regional power systems, i.e., a polytope in the domain of ($\mathbf{w}_1$, …, $\mathbf{w}_t$, …, $\mathbf{w}_{n_T}$) where $\mathbf{w}_t$ is the vector of the coupling parameters at time period $t$ and $n_T$ is the total number of consecutive periods. For each point within the tie-line security region, at least one feasible decision of the regional power system exists. This is defined as the feasibility of the tie-line security region in this paper.

The dimension of a tie-line security region is $\sum_{t=1}^{T} n_{wt}$, where $n_{wt}$ is the dimension of $\mathbf{w}_t$. Generally, multiple time periods are considered in power system operations and multiple border buses are inevitable in utility systems. This casts the tie-line security region as a high-dimension polytope. The computational complexity of exploring a polytope fast increase with the dimension of the explored polytope, as also found in [7]-[8].

The representatives to calculate a tie-line security region include Fuourier-Motzkin elimination [9]-[11], multi-parametric programming [12]-[13] and vertex search [14]-[15]. The tie-line security regions in [9]-[14] considered a single period and only a few border buses. When these methods are extended to the multiple time periods and more border buses, a considerable computational burden would suffer, as will be detailedly reviewed in Sec. II. Consequently, the fast calculation of the tie-line security region considering time coupling was studied in [15]. As reported in [15], the original high-dimension polytope that represents the tie-line security region with time coupling was simply decomposed as the union of a few lower-dimension polytopes associated with combinations of different periods. The computational burden brought by multiple time periods was alleviated while the computational burden brought by multiple border buses remained. Also, the feasibility of the tie-line security region in [15] cannot be guaranteed via the simple decomposition. At the same time, the information of renewable accommodations over the tie-line security region remained unknown in [15]. This leads to the difficulty of coordinating regional power systems via tie-line security regions to facilitate renewable accommodations accommodate. These limitations in [15] mentioned above will be further elaborated in Sec. II.

To overcome the defects mentioned above, the fast calculation of a tie-line security region in high dimension for renewable accommodations is studied in this paper. Two works are contained in this paper, as described below.

W. Lin is with the Department of Information Engineering, The Chinese University of Hong Kong, Hong Kong, SAR, China. H. Jiang and Z. Yang are with the State Key Laboratory of Power Transmission Equipment & System Security and New Technology, College of Electrical Engineering, Chongqing University, Chongqing, China. (*Corresponding author: W. Lin, wlin@ie.cuhk.edu.hk*)

(1) A fast calculation method of a tie-line security region that considers multiple time periods and border buses is presented (Sec. IV). The computational burden brought by multiple time periods is avoided by decomposing the original tie-line security region across all the time periods as a Cartesian production of lower-dimension polytopes at each time period. This decomposition is achieved by leveraging dispatch levels of generations. Furthermore, the computational burden brought by multiple border buses is alleviated by aggregating tie-line power. This paves a promising way to represent the high-dimension polytope in a lower dimension. Proofs are provided for the legibility of our method with which the feasibility of our tie-line security region is guaranteed. Also, the minimum renewable curtailments of each regional power system are preserved over the tie-line security region by incorporating an additional dimension in the tie-line security region.

(2) The application method of the tie-line security region in renewable accommodations under a decentralized and non-iterative framework is illustrated (Sec. V). The coupling constraints among regional power systems are reformulated based on the vertices of tie-line security regions. Given the reformulated coupling constraints, a *linear programming* (LP) problem to reduce renewable curtailments in an interconnected power system is established by coordinating the coupling parameters.

## II. LITERATURE REVIEW

The representatives [9]-[15] to describe the power transfer capability as a tie-line security region stem from the following optimization problem of a regional power system:

$$\min_{\mathbf{x}} f(\mathbf{x}), \quad (1)$$

$$\text{s.t.} \quad \mathbf{Ax} + \mathbf{Bw} \leq \mathbf{C}, \quad (2)$$

where $\mathbf{x}$ is the vector of decision variables; $\mathbf{w}$ is the vector of the coupling parameters among different regional power systems; $f(\mathbf{x})$ is a convex objective function with respect to $\mathbf{x}$ (e.g., renewable curtailments, and fuel costs); $\mathbf{A}$ and $\mathbf{B}$ are constant coefficient matrices, and $\mathbf{C}$ is a constant coefficient vector. Note that the DC power flow model has been widely adopted in power industries due to its robustness and efficiency [18]. This leads to the operational requirements formulated via the DC power flow model, i.e., the linear constraint (2). Under such a case, the coupling parameters will include tie-line power and border angles in this paper.

In representatives, tie-line security regions are represented in the domain of $\mathbf{w}$ with which there is a feasible $\mathbf{x}$ that satisfies the constraint (2). Based on different ideas, they can be categorized into three types.

*Type 1*: *methods based on Fuourier-Motzkin elimination*. A tie-line security region can be regarded as the projection from the domain of $(\mathbf{x},\mathbf{w})$ to the domain of $\mathbf{w}$. This projection can be implemented by Fourier-Motzkin elimination. This similar idea has been employed to project the generation-demand space onto the demand space [9]-[10]. In [11], Fuourier-Motzkin elimination was employed to get the feasible region of power exchanges at the point of common coupling in distribution networks[11]. In Fuourier-Motzkin elimination, a variable can be bounded by a linear combination of other variables. This provides the opportunity to eliminate the unfocused variables iteratively. However, each variable is eliminated at the expense of adding numerous inequalities [16]. The computational burden exponentially increases with the number of eliminated variables. Consequently, Fuourier-Motzkin elimination may suffer the heavy computational burden when multiple time periods are considered. Also, Fuourier-Motzkin elimination only calculates the tie-line security region, while the information of the objective function of a regional power system remains unknown.

*Type 2*: *methods based on multi-parametric programming*. Multi-parametric programing was employed in [12]-[13] to calculate a single-period tie-line security region by regarding the coupling parameters as programming parameters. As reported in [12], each combination of active and inactive constraints of (1)-(2), which can be obtained by repeatedly solving the optimization problem in (1)-(2), describes a subregion of the tie-line security region. Consequently, an exact tie-line security region can be obtained by the exhaustive enumeration of all the combinations of active and inactive constraints. Also, the optimal objective function can be formulated as a piecewise affine function of the coupling parameters. To alleviate the computational burden arising from the exhaustive enumeration, a modified multi-parametric programming method was reported in [13]. The exact tie-line security region in [13] was calculated by only enumerating combinations of active and inactive constraints that are related to boundaries. However, the computational burden of multi-parametric programming fast increases with the number of both programing parameters and constraints [17]. Multiple time periods and multiple border buses will increase the number of both programing parameters and constraints. Consequently, the computational burden of multi-parametric programming will become heavy when multiple time periods and multiple border buses are involved.

*Type 3*: *methods based on vertex search*. Different with Fourier-Motzkin elimination, the projection to calculate the tie-line security region was achieved by exploring vertices of the tie-line security region [14]-[15]. For example, vertices in [14] were obtained by solving a series of LP problems to construct a single-period tie-line security region. However, the number of vertices fast increases with the dimension of a tie-line security region. To overcome the computational burden brought by multiple time periods, the fast calculation of a multi-period tie-line security region was studied in [15]. The original high-dimension tie-line security region was simply decomposed as the union of a few lower-dimension polytopes. In each lower-dimension polytopes, the tie-line security region was projected to the domain of coupling parameters at certain time periods. Consequently, the method in [15] was weak to handle multiple border buses. Also, the decomposition in [15] is heuristic. This leads to the absence of guaranteeing the feasibility of the tie-line security region. Note that vertex search in [14] and [15] only calculates the tie-line security region, while the information of the objective function is unknown.

As mentioned above, the representatives [9]-[15] still face troubles to calculate the tie-line security region and its information of the objective function, when multiple time periods and multiple border buses inherently in power system operations are involved. This motivates us to fill this research gap in this paper.

III. PROBLEM FORMULATION

The optimization problem to minimize renewable curtailments in an interconnected power system will be presented in this section. The model is constructed based on the DC power flow model that has been widely used in power industries [18].

A. *The operating region $X_{q,t}$ at time period $t$ in the regional power system $q$*

(1) Power balance should equal power demand, i.e.,

$$\mathbf{e}_{G,q,t}^T \mathbf{P}_{G,q,t} + \mathbf{e}_{B,q,t}^T \mathbf{P}_{B,q,t} + \mathbf{e}_{R,q,t}^T \left( \mathbf{P}_{R,q,t} - \mathbf{C}_{R,q,t} \right) = \mathbf{e}_{D,q,t}^T \mathbf{P}_{D,q,t}, \quad (3)$$

where $\mathbf{P}_{G,q,t}$, $\mathbf{P}_{B,q,t}$, $\mathbf{P}_{R,q,t}$, $\mathbf{C}_{R,q,t}$, and $\mathbf{P}_{D,q,t}$ represent the vectors of the generation levels, tie-line power, maximum renewable power, renewable curtailments, and power demand at time period $t$ in the regional power system $q$, respectively; $\mathbf{e}_{G,q,t}$, $\mathbf{e}_{B,q,t}$, $\mathbf{e}_{R,q,t}$, and $\mathbf{e}_{D,q,t}$ are all-one vectors whose dimensions correspondingly match $\mathbf{P}_{G,q,t}$, $\mathbf{P}_{B,q,t}$, $\mathbf{P}_{R,q,t}$, and $\mathbf{P}_{D,q,t}$.

(2) Relationship between border angles and power injections can be described below.

$$\boldsymbol{\theta}_{B,q,t} = \mathbf{B}_{G,q,t} \mathbf{P}_{G,q,t} + \mathbf{B}_{B,q,t} \mathbf{P}_{B,q,t} + \mathbf{B}_{R,q,t} \left( \mathbf{P}_{R,q,t} - \mathbf{C}_{R,q,t} \right) + \mathbf{B}_{D,q,t} \mathbf{P}_{D,q,t}, \quad (4)$$

where $\boldsymbol{\theta}_{B,q,t}$ is the vector of the border angles at time period $t$ in the regional power system $q$; $\mathbf{B}_{G,q,t}=\mathbf{Y}_{q,t}\mathbf{M}_{G,q,t}$; $\mathbf{B}_{B,q,t}=\mathbf{Y}_{q,t}\mathbf{M}_{B,q,t}$; $\mathbf{B}_{R,q,t}=\mathbf{Y}_{q,t}\mathbf{M}_{R,q,t}$; $\mathbf{B}_{D,q,t}=\mathbf{Y}_{q,t}\mathbf{M}_{D,q,t}$; $\mathbf{Y}_{q,t}$ is the inverse matrix of the susceptance matrix associated with border buses at time period $t$ in the regional power system $q$; $\mathbf{M}_{G,q,t}$, $\mathbf{M}_{B,q,t}$, $\mathbf{M}_{R,q,t}$, and $\mathbf{M}_{D,q,t}$ are incident matrices associated with $\mathbf{P}_{G,q,t}$, $\mathbf{P}_{B,q,t}$, $\mathbf{P}_{R,q,t}$, and $\mathbf{P}_{D,q,t}$, respectively.

(3) Branch flows should not exceed limits, i.e.,

$$\mathbf{A}_{G,q,t} \mathbf{P}_{G,q,t} + \mathbf{A}_{B,q,t} \mathbf{P}_{B,q,t} + \mathbf{A}_{R,q,t} \left( \mathbf{P}_{R,q,t} - \mathbf{C}_{R,q,t} \right) + \mathbf{A}_{D,q,t} \mathbf{P}_{D,q,t} \leq \mathbf{P}_{F,q,t}^{\max}, \quad (5)$$

$$\mathbf{P}_{F,q,t}^{\min} \leq \mathbf{A}_{G,q,t} \mathbf{P}_{G,q,t} + \mathbf{A}_{B,q,t} \mathbf{P}_{B,q,t} + \mathbf{A}_{R,q,t} \left( \mathbf{P}_{R,q,t} - \mathbf{C}_{R,q,t} \right) + \mathbf{A}_{D,q,t} \mathbf{P}_{D,q,t}, \quad (6)$$

where $\mathbf{P}_{F,q,t}$ is the vector of the branch flows at time period $t$ in the regional power system $q$; $\mathbf{P}_{F,q,t}^{\min}$ and $\mathbf{P}_{F,q,t}^{\max}$ are the lower and upper bounds of $\mathbf{P}_{F,q,t}$, respectively; $\mathbf{A}_{G,q,t}=\mathbf{S}_{q,t}\mathbf{M}_{G,q,t}$; $\mathbf{A}_{B,q,t}=\mathbf{S}_{q,t}\mathbf{M}_{B,q,t}$; $\mathbf{A}_{R,q,t}=\mathbf{S}_{q,t}\mathbf{M}_{R,q,t}$; $\mathbf{A}_{D,q,t}=\mathbf{S}_{q,t}\mathbf{M}_{D,q,t}$; $\mathbf{S}_{q,t}$ is the *power transfer distribution factor* (PTDF) matrix at time period $t$ in the regional power system $q$.

(4) Generation levels should not exceed dispatch levels, i.e.,

$$\mathbf{P}_{G,q,t}^{\min} \leq \mathbf{P}_{G,q,t} \leq \mathbf{P}_{G,q,t}^{\max}, \quad (7)$$

where $\mathbf{P}_{G,q,t}^{\min}$ and $\mathbf{P}_{G,q,t}^{\max}$ are the vectors of the dispatch levels of $\mathbf{P}_{G,q,t}$. $\mathbf{P}_{G,q,t}^{\min}$ and $\mathbf{P}_{G,q,t}^{\max}$ will be selected in Sec. III-B.

(5) Tie-line power should not exceed its flow limits, i.e.,

$$\mathbf{P}_{B,q,t}^{\min} \leq \mathbf{P}_{B,q,t} \leq \mathbf{P}_{B,q,t}^{\max}, \quad (8)$$

where $\mathbf{P}_{B,q,t}^{\min}$ and $\mathbf{P}_{B,q,t}^{\max}$ are lower and upper limits, respectively.

(6) Renewable curtailments should not exceed the maximum renewable power, i.e.,

$$\mathbf{0} \leq \mathbf{C}_{R,q,t} \leq \mathbf{P}_{R,q,t}. \quad (9)$$

(7) Border angles should be located between $-\pi$ and $\pi$, i.e.,

$$-\boldsymbol{\pi} \leq \boldsymbol{\theta}_{B,q,t} \leq \boldsymbol{\pi}. \quad (10)$$

(8) The epigraph of total renewable curtailments with an upper bound at time period $t$, i.e.,

$$\mathbf{e}_{R,q,t}^T \mathbf{C}_{R,q,t} \leq z_{q,t} \leq \mathbf{e}_{R,q,t}^T \mathbf{P}_{R,q,t}, \forall q, \forall t, \quad (11)$$

where $z_{q,t}$ is a continuous variable that represents the total renewable curtailments at time period $t$ in the regional power system $q$. This constraint can be combined with the coming objective function (14) for the optimal dispatch to reduce renewable curtailments, as similarly done in [19].

(9) Generation level changes between two consecutive time periods should not exceed ramp rates, i.e.,

$$\mathbf{R}_{G,q,t}^{\text{down}} \leq \mathbf{P}_{G,q,t} - \mathbf{P}_{G,q,t-1} \leq \mathbf{R}_{G,q,t}^{\text{up}}, \quad (12)$$

where $\mathbf{R}_{G,q,t}^{\text{down}}$ and $\mathbf{R}_{G,q,t}^{\text{up}}$ are the vectors of ramp-down and ramp-up rates, respectively.

B. *Coupling between the regional power system $q$ and the regional power system $p$*

Coupling between two regional power systems is reflected in the electronic relationship described below.

$$\mathbf{P}_{B,q,t} = \mathbf{P}_{B,p,t} = \mathbf{D}_{B,p,t} \boldsymbol{\theta}_{B,p,t} + \mathbf{D}_{B,q,t} \boldsymbol{\theta}_{B,q,t}, \quad (13)$$

where $\mathbf{D}_{B,p,t}=\mathbf{X}_{q,p,t}\mathbf{A}_{B,p,t}$; $\mathbf{D}_{B,q,t}=\mathbf{X}_{q,p,t}\mathbf{A}_{B,q,t}$; $\mathbf{X}_{q,p,t}$ is the resistance matrix at time period $t$ between the regional power system $q$ and the regional power system $p$; $\mathbf{A}_{B,p,t}$ and $\mathbf{A}_{B,q,t}$ are incident matrices associated with $\boldsymbol{\theta}_{B,p,t}$ and $\boldsymbol{\theta}_{B,q,t}$, respectively.

For illustration purposes, each border bus is only connected with a tie-line in this paper.

C. *The objective function and the compact formulation of the optimization problem*

The goal in an interconnected power system is to minimize renewable curtailments in this paper, i.e.,

$$\min_{\{z_{q,t}, \mathbf{P}_{G,q,t}, \mathbf{C}_{R,q,t}\}, \forall t, \forall q} \sum_{q=1}^{n_Q} \sum_{t=1}^{n_T} z_{q,t}, \quad (14)$$

where $n_Q$ is the number of regional power systems; $n_T$ is the number of time periods.

The optimization problem to minimize renewable curtailments in an interconnected power system is defined as the OP1 problem below.

**OP1**: The objective function (14) subjects to the constraints (3)-(13).

For the OP1 problem, a decision vector at time period $t$ in the regional power system $q$ is denoted as $\mathbf{x}_{q,t}=\begin{bmatrix}\mathbf{P}_{G,q,t}^T & \mathbf{C}_{R,q,t}^T\end{bmatrix}^T$ and we have $\mathbf{x}_q=\begin{bmatrix}\mathbf{x}_{q,1}^T & \cdots & \mathbf{x}_{q,n_T}^T\end{bmatrix}^T$. A vector $\mathbf{z}_q$ is denoted as $\mathbf{z}_q=\begin{bmatrix}z_{q,1} & \cdots & z_{q,n_T}\end{bmatrix}^T$. The operating region across all the time periods in the regional power system $q$ is denoted as $\mathbf{X}_q=\mathbf{X}_{q,1} \times \ldots \times \mathbf{X}_{q,n_T}$. The tie-line power associated with the regional power system $q$ is denoted as $\mathbf{P}_{B,q}=\begin{bmatrix}\mathbf{P}_{B,q,1}^T & \cdots & \mathbf{P}_{B,q,n_T}^T\end{bmatrix}^T$. The border angles associated with the regional power system $q$ is denoted as $\boldsymbol{\theta}_{B,q}=[\boldsymbol{\theta}_{B,q,1}^T \; ... \; \boldsymbol{\theta}_{B,q,n_T}^T]^T$. For convenience of discussion, a compact form of the OP1 problem is given below.

$$\min_{\{\mathbf{z}_q,\mathbf{x}_q\},\forall t,\forall q} \sum_{q=1}^{n_O} \mathbf{e}_q^T \mathbf{z}_q, \quad (15)$$

$$\text{s.t.} \quad \text{Constraint (13)}, \quad (16)$$

$$(\mathbf{x}_q,\mathbf{P}_{B,q},\boldsymbol{\theta}_{B,q},\mathbf{z}_q) \in \mathbf{X}_q. \quad (17)$$

The objective function (15) is the compact form of the objective function (14), where $\mathbf{e}_q$ is an all-one vector associated with $\mathbf{z}_q$. The constraint (16) is exactly the constraint (13). The constraint (17) is the compact form of the constraints (3)-(12). As observed in (15)-(17), if the coupling parameters $\mathbf{P}_{B,q}$ and $\boldsymbol{\theta}_{B,q}$ are fixed, the decisions $\mathbf{x}_q$ of each regional power system can be separately determined. This goal will be achieved in this paper by calculating the tie-line security region in the domain of $(\mathbf{P}_{B,q},\mathbf{z}_q)$. The tie-line security region $\boldsymbol{\Omega}_q$ in the regional power system $q$ is mathematically defined as below.

$$\boldsymbol{\Omega}_q \triangleq \left\{ (\mathbf{P}_{B,q},\mathbf{z}_q) \mid \exists (\mathbf{x}_q,\boldsymbol{\theta}_{B,q}) \in \mathbf{X}_q \right\}. \quad (18)$$

**Remark 1.** A tie-line security region is traditionally discussed in the domain of the coupling parameters (i.e., $\mathbf{P}_{B,q}$ and $\boldsymbol{\theta}_{B,q}$), as reported in [12]-[13]. In this paper, the domain of the tie-line security region is slightly different with the traditional one because the tie-line security region is discussed in the domain of $(\mathbf{P}_{B,q},\mathbf{z}_q)$. This will bring two advantages.

1) The incorporation of $\mathbf{z}_q$ can preserve the information of the objective function over the tie-line security region, as will be further elaborated in Remark 4.

2) The dimension of the traditional tie-line security region decreases at the absence of $\boldsymbol{\theta}_{B,q}$. Also, this neglection will not affect the coordination among regional power systems because $\boldsymbol{\theta}_{B,q}$ can be explicity formulated as a function of $(\mathbf{P}_{B,q},\mathbf{z}_q)$ based on via our application method in Sec. V. ∎

If the number of border buses is denoted as $n_B$, the dimension of $\boldsymbol{\Omega}_q$ will be $n_T \times n_{B,q}+n_T$, where $n_{B,q}$ is the number of border buses in the regional power system $q$. The dimension of $\boldsymbol{\Omega}_q$ is high when multiple time periods and multiple border buses inherently in power system operations are involved. This leads to the heavy computational burden. To this end, the computational burden of calculating the high-dimension tie-line security region $\boldsymbol{\Omega}_q$ will be alleviated in Sec. IV.

## IV. FAST CALCULATION OF A TIE-LINE SECURITY REGION IN HIGH DIMENSION

The computational burden of calculating $\boldsymbol{\Omega}_q$ arises from time periods and border buses. Consequently, our method presented in this section alleviates the computational burden by decomposing time periods and aggregating tie-line power, as will be correspondingly elaborated in Sec. IV-A and Sec. IV-B.

### A. Decomposition of time periods

The key issue of decomposing time periods lies in selecting appropriate dispatch levels (i.e., $\mathbf{P}_{G,q,t}^{\min}$ and $\mathbf{P}_{G,q,t}^{\max}$ in (7)) with which the ramp rate constraint (12) can be neglected while this constraint can be naturally fulfilled. This goal can be achieved by solving the following LP problem.

$$\max_{\{z_{q,t},\mathbf{x}_{q,t},\mathbf{P}_{B,q,t},\boldsymbol{\theta}_{B,q,t},\mathbf{P}_{G,q,t}^{\max},\mathbf{P}_{G,q,t}^{\min}\},\forall t} \sum_{t=1}^{n_T} \mathbf{e}_{G,q,t}^T \left(\mathbf{P}_{G,q,t}^{\max}-\mathbf{P}_{G,q,t}^{\min}\right), \quad (19)$$

$$\text{s.t.} \quad \text{Constraints (3)-(11)}, \forall t, \quad (20)$$

$$\mathbf{P}_{G,q,t}^{\max} - \mathbf{P}_{G,q,t-1}^{\min} \leq \mathbf{R}_{G,t}^{\text{up}}, \forall t, \quad (21)$$

$$\mathbf{P}_{G,q,t-1}^{\max} - \mathbf{P}_{G,q,t}^{\min} \leq -\mathbf{R}_{G,q,t}^{\text{down}}, \forall t, \quad (22)$$

$$\underline{\mathbf{P}}_{G,q} \leq \mathbf{P}_{G,q,t}^{\min} \leq \mathbf{P}_{G,q,t}^{\max} \leq \overline{\mathbf{P}}_{G,q}, \forall t, \quad (23)$$

where $\overline{\mathbf{P}}_{G,q}$ and $\underline{\mathbf{P}}_{G,q}$ are the vectors of generator capacities in the regional power system $q$.

When the dispatch levels $\mathbf{P}_{G,q,t}^{\min}$ and $\mathbf{P}_{G,q,t}^{\max}$ are determined by solving the LP problem (19)-(23), the following Proposition 1 shows that the ramp rate constraint (12) can be fulfilled.

**Proposition 1.** Define: 1) $\mathbf{Y}_{q,t}$ as the operating region where $\mathbf{X}_{q,t}$ excludes the ramp rate constraint (12), and 2) $\mathbf{Y}_q=\mathbf{Y}_{q,1}\times\ldots\times\mathbf{Y}_{q,n_T}$. For $\mathbf{Y}_{q,t}$, a tie-line security region $\boldsymbol{\Omega}_{q,t}$ is defined as

$$\boldsymbol{\Omega}_{q,t} \triangleq \left\{ (\mathbf{P}_{B,q,t},z_{q,t}) \mid \exists (\mathbf{x}_{q,t},\boldsymbol{\theta}_{B,q,t}) \in \mathbf{Y}_{q,t} \right\}. \quad (24)$$

The dispatch levels $\mathbf{P}_{G,q,t}^{\min}$ and $\mathbf{P}_{G,q,t}^{\max}$ determined from the LP problem (19)-(23) yield

$$\boldsymbol{\Omega}_q = \mathbf{R}_q, \quad (25)$$

where $\mathbf{R}_q$ is defined as

$$\mathbf{R}_q = \boldsymbol{\Omega}_{q,1} \times \cdots \times \boldsymbol{\Omega}_{q,t} \times \cdots \times \boldsymbol{\Omega}_{q,n_T}. \quad (26)$$
∎

**Proof.** This proof is completed by two steps: 1) $\mathbf{R}_q$ is a subset of $\boldsymbol{\Omega}_q$, and 2) $\boldsymbol{\Omega}_q$ is a subset of $\mathbf{R}_q$.

1) $\mathbf{R}_q$ is a subset of $\boldsymbol{\Omega}_q$. For the point $(\mathbf{P}_{B,q},\mathbf{z}_q)$ within $\mathbf{R}_q$, a feasible $(\mathbf{x}_q, \boldsymbol{\theta}_{B,q})$ exists within $\mathbf{Y}_q$. Note that $\mathbf{X}_q$ contains the constraints (3)-(12) across all the time periods while $\mathbf{Y}_q$ only contains the constraints (3)-(11) across all the time periods. Consequently, the constraints (3)-(11) in $\mathbf{X}_q$ must be fulfilled with the same $(\mathbf{x}_q,\boldsymbol{\theta}_{B,q})$. Our major focus comes down to the constraint (12) in $\mathbf{X}_q$. Its fulfillment can be proven as below.

The constraint (7) yields

$$\mathbf{P}_{G,q,t}^{\min} - \mathbf{P}_{G,q,t-1}^{\max} \leq \mathbf{P}_{G,q,t} - \mathbf{P}_{G,q,t-1} \leq \mathbf{P}_{G,q,t}^{\max} - \mathbf{P}_{G,q,t-1}^{\min}. \quad (27)$$

If the dispatch levels $\mathbf{P}_{G,q,t}^{\min}$ and $\mathbf{P}_{G,q,t}^{\max}$ are determined from the LP problem (19)-(23), (27) can be further expressed as follows:

$$\mathbf{R}_{G,q,t}^{\text{down}} \leq \mathbf{P}_{G,q,t}^{\min} - \mathbf{P}_{G,q,t-1}^{\max} \leq \mathbf{P}_{G,q,t} - \mathbf{P}_{G,q,t-1} \leq \mathbf{P}_{G,q,t}^{\max} - \mathbf{P}_{G,q,t-1}^{\min} \leq \mathbf{R}_{G,q,t}^{\text{up}}. \quad (28)$$

Based on (28), $(\mathbf{x}_q,\boldsymbol{\theta}_{B,q})$ within $\mathbf{Y}_q$ also satisfies the constraint (12) in $\mathbf{X}_q$. Consequently, the point $(\mathbf{P}_{B,q}, \mathbf{z}_q)$ within $\mathbf{R}_q$ is also within $\boldsymbol{\Omega}_q$, i.e., $\mathbf{R}_q$ is a subset of $\boldsymbol{\Omega}_q$.

2) $\boldsymbol{\Omega}_q$ is a subset of $\mathbf{R}_q$. For the point $(\mathbf{P}_{B,q}, \mathbf{z}_q)$ within $\boldsymbol{\Omega}_q$, a feasible $(\mathbf{x}_q, \boldsymbol{\theta}_{B,q})$ exists within $\mathbf{X}_q$. Note that $\mathbf{X}_q$ contains the constraints (3)-(12) across all the time periods while $\mathbf{Y}_q$ only contains the constraints (3)-(11) across all the time periods. Consequently, for the same point $(\mathbf{P}_{B,q}, \mathbf{z}_q)$, the same $(\mathbf{x}_q, \boldsymbol{\theta}_{B,q})$ within $\mathbf{X}_q$ is also within $\mathbf{Y}_q$. In other words, the point $(\mathbf{P}_{B,q}, \mathbf{z}_q)$ within $\boldsymbol{\Omega}_q$ is also within $\mathbf{R}_q$, i.e., $\mathbf{R}_q$ is a subset of $\boldsymbol{\Omega}_q$. ∎

**Remark 2.** Proposition 1 implies that the complete tie-line security region across all the time periods can be a Cartesian production of tie-line security regions at each time period. Consequently, the calculation of $\boldsymbol{\Omega}_q$ relies on the calculation of $\Omega_{q,t}$ that is in the domain of ($\mathbf{P}_{B,q,t}$, $z_{q,t}$). This alleviates the computational burden brought by multiple time periods. ∎

As demonstrated in Proposition 1 and Remark 2, the tie-line security region $\Omega_{q,t}$ should be calculated. However, the dimension of $\Omega_{q,t}$ is $n_{B,q}+1$. When facing multiple border buses, the computational burden of calculating $\Omega_{q,t}$ would be heavy. This will be resolved in next subsection by aggregating tie-line power.

*B. Aggregation of tie-line power*

The key idea of reducing the dimension of $\Omega_{q,t}$ lies in representing $\Omega_{q,t}$ in a new domain of ($\widetilde{\mathbf{P}}_{B,q,t}$, $z_{q,t}$) instead of the original domain of ($\mathbf{P}_{B,q,t}$, $z_{q,t}$), where the dimension of $\widetilde{\mathbf{P}}_{B,q,t}$ is smaller than that of $\mathbf{P}_{B,q,t}$. This can be achieved by establishing a mapping function from $\mathbf{P}_{B,q,t}$ to $\widetilde{\mathbf{P}}_{B,q,t}$. In this paper, the mapping function is explicitly established by 1) categorizing tie-line power into groups, and 2) aggregating $\mathbf{P}_{B,q,t}$ as $\widetilde{\mathbf{P}}_{B,q,t}$, as will be elaborated in coming contents.

Geographically, different tie-lines are located between different regional power systems. In this paper, $\mathbf{P}_{B,q,t}$ is categorized based on its geographic locations, i.e., the tie-line power at the same border interface is categorized into the same group. For the tie-line power in each group, they are aggregated as one equivalent tie-line power as described below.

$$\tilde{P}_{B,q,t}^{j} = \sum_{i=1}^{n_j} P_{B,q,t}^{j_i}, \forall j, \quad (29)$$

where $P_{B,q,t}^{j_i}$ is the tie-line power from $\mathbf{P}_{B,q,t}$ and belongs to the $j^{th}$ group; $\tilde{P}_{B,q,t}^{j}$ is the aggregated tie-line power in the $j^{th}$ group; $n_j$ is the number of aggregated tie-lines in the $j^{th}$ group.

Based on the aggregation in (29), $\mathbf{P}_{B,q,t}$ can be mapped to $\widetilde{\mathbf{P}}_{B,q,t}=[\tilde{P}_{B,q,t}^{1} \ ... \ \tilde{P}_{B,q,t}^{K}]^T$. This provides an opportunity to represent $\Omega_{q,t}$ in the domain of ($\widetilde{\mathbf{P}}_{B,q,t}$, $z_{q,t}$) instead of ($\mathbf{P}_{B,q,t}$, $z_{q,t}$). However, $\Omega_{q,t}$ defined in (24) arises from the operating region $\mathbf{Y}_{q,t}$ where $\mathbf{P}_{B,q,t}$ instead of $\widetilde{\mathbf{P}}_{B,q,t}$ exists. Proposition 2 is presented to incorporate $\widetilde{\mathbf{P}}_{B,q,t}$ into $\mathbf{Y}_{q,t}$.

**Proposition 2.** Denote a polytope $\widetilde{\Omega}_{q,t}$ as

$$\widetilde{\Omega}_{q,t} \triangleq \left\{ (\widetilde{\mathbf{P}}_{B,q,t}, z_{q,t}) \middle| \exists (\mathbf{x}_{q,t}, \boldsymbol{\theta}_{B,q,t}, \mathbf{P}_{B,q,t}) \in \widetilde{\mathbf{Y}}_{q,t} \right\}, \quad (30)$$

where $\widetilde{\mathbf{Y}}_{q,t}$ is revised from $\mathbf{Y}_{q,t}$ as follows:

$$\widetilde{\mathbf{P}}_{B,q,t} = \widetilde{\mathbf{A}}_{B,q,t}\mathbf{P}_{B,q,t}, \quad (31)$$

$$\mathbf{e}_{G,q,t}^T\mathbf{P}_{G,q,t} + \tilde{\mathbf{e}}_{B,q,t}^T\widetilde{\mathbf{P}}_{B,q,t} + \mathbf{e}_{R,q,t}^T(\mathbf{P}_{R,q,t}-\mathbf{C}_{R,q,t}) = \mathbf{e}_{D,q,t}^T\mathbf{P}_{D,q,t}, (32)$$

$$\begin{aligned}&\mathbf{A}_{G,q,t}\mathbf{P}_{G,q,t} + \left(\tilde{\boldsymbol{\alpha}}_{B,q,t}\widetilde{\mathbf{P}}_{B,q,t} + \tilde{\boldsymbol{\beta}}_{B,q,t}\right)\\&+\mathbf{A}_{R,q,t}(\mathbf{P}_{R,q,t}-\mathbf{C}_{R,q,t}) + \mathbf{A}_{D,q,t}\mathbf{P}_{D,q,t} \le \mathbf{P}_{F,q,t}^{\max} - \boldsymbol{\varepsilon}_{B,q,t},\end{aligned} \quad (33)$$

$$\begin{aligned}&\mathbf{P}_{F,q,t}^{\min} + \boldsymbol{\varepsilon}_{B,q,t} \le \mathbf{A}_{G,q,t}\mathbf{P}_{G,q,t} + \left(\tilde{\boldsymbol{\alpha}}_{B,q,t}\widetilde{\mathbf{P}}_{B,q,t} + \tilde{\boldsymbol{\beta}}_{B,q,t}\right)\\&+\mathbf{A}_{R,q,t}(\mathbf{P}_{R,q,t}-\mathbf{C}_{R,q,t}) + \mathbf{A}_{D,q,t}\mathbf{P}_{D,q,t},\end{aligned} \quad (34)$$

$$\mathbf{P}_{B,q,t}^{\min,re} \le \mathbf{P}_{B,q,t} \le \mathbf{P}_{B,q,t}^{\max,re}, \quad (35)$$

$$\text{Constraints (4), (7) and (9)-(11)}. \quad (36)$$

The notations mentioned above can be found in the constraints (3)-(11), except: 1) the constant matrix $\widetilde{\mathbf{A}}_{B,q,t}$ in the constraint (31), 2) the vectors $\mathbf{P}_{B,q,t}^{\min,re}$ and $\mathbf{P}_{B,q,t}^{\max,re}$ in the constraint (35), and 3) the vectors $\tilde{\boldsymbol{\alpha}}_{B,q,t}$ and $\tilde{\boldsymbol{\beta}}_{B,q,t}$ in the constraints (33)-(34), and 4) the error bound $\boldsymbol{\varepsilon}_{B,q,t}$ in the constraints (33)-(34). They will be elaborated below.

The constraint (31) is a compact form of (29), leading to the easy calculation of $\widetilde{\mathbf{A}}_{B,q,t}$ based on (29). Each element in $\mathbf{P}_{B,q,t}^{\min,re}$ (resp. $\mathbf{P}_{B,q,t}^{\max,re}$) can be determined by solving an LP problem where the objective function is the minimum (resp. maximum) corresponding tie-line power subjected to the constraints (3)-(11). For $\tilde{\boldsymbol{\alpha}}_{B,q,t}$ and $\tilde{\boldsymbol{\beta}}_{B,q,t}$, they are determined based on the method presented in Appendix. As presented in Appendix, the following property holds:

$$-\boldsymbol{\varepsilon}_{B,q,t} \le \mathbf{A}_{B,q,t}\mathbf{P}_{B,q,t} - \left(\tilde{\boldsymbol{\alpha}}_{B,q,t}\widetilde{\mathbf{P}}_{B,q,t} + \tilde{\boldsymbol{\beta}}_{B,q,t}\right) \le \boldsymbol{\varepsilon}_{B,q,t}. \quad (37)$$

Note that the error bound $\boldsymbol{\varepsilon}_{B,q,t}$ is non-negative and is an explicit function of $\tilde{\boldsymbol{\alpha}}_{B,q,t}$ and $\tilde{\boldsymbol{\beta}}_{B,q,t}$, as also calculated in Appendix.

Based on definitions mentioned above, for the point ($\widetilde{\mathbf{P}}_{B,q,t}$, $z_{q,t}$) within $\widetilde{\Omega}_{q,t}$, a feasible ($\mathbf{x}_{q,t}$, $\boldsymbol{\theta}_{B,q,t}$, $\mathbf{P}_{B,q,t}$) exists within $\widetilde{\mathbf{Y}}_{q,t}$. At the same time, ($\mathbf{x}_{q,t}$, $\boldsymbol{\theta}_{B,q,t}$, $\mathbf{P}_{B,q,t}$) within $\widetilde{\mathbf{Y}}_{q,t}$ also satisfies the constraints (3)-(11) that construct $\mathbf{Y}_{q,t}$. In other words, $\widetilde{\Omega}_{q,t}$ provides an inner estimation of $\Omega_{q,t}$ in the domain of ($\widetilde{\mathbf{P}}_{B,q,t}$, $z_{q,t}$). ∎

**Proof.** This proof is completed by checking the existence of the constraints (3)-(11) when ($\mathbf{x}_{q,t}$, $\boldsymbol{\theta}_{B,q,t}$, $\mathbf{P}_{B,q,t}$) within $\widetilde{\mathbf{Y}}_{q,t}$ is given.

➤ The constraints (4) and (9)-(11) hold because they are also incorporated in $\widetilde{\mathbf{Y}}_{q,t}$ via the constraint (36).

The existence of the constraint (3) can be checked based on the constraints (31)-(32). The constraint (31) is a compact form of (29) that describe the aggregation of tie-line power. Particularly, each tie-line power only belongs to a group. This feature yields

$$\tilde{\mathbf{e}}_{B,q,t}^T\widetilde{\mathbf{P}}_{B,q,t} = \mathbf{e}_{B,q,t}^T\mathbf{P}_{B,q,t}, \quad (38)$$

where $\tilde{\mathbf{e}}_{B,q,t}$ is an all-one vector associated with $\widetilde{\mathbf{P}}_{B,q,t}$.

Substituting (38) into the constraint (3) yields the constraint (32), i.e., the constraint (3) holds.

➤ The constraint (8) holds because the parameters in (35) are determined when the constraint (8) holds.

➤ The constraints (5)-(6) hold due to the constraints (33), (34) and (37). The constraint (37) is expanded as follows:

$$\mathbf{A}_{B,q,t}\mathbf{P}_{B,q,t} - \boldsymbol{\varepsilon}_{B,q,t} \le \tilde{\boldsymbol{\alpha}}_{B,q,t}\widetilde{\mathbf{P}}_{B,q,t} + \tilde{\boldsymbol{\beta}}_{B,q,t}, \quad (39)$$

$$\tilde{\boldsymbol{\alpha}}_{B,q,t}\widetilde{\mathbf{P}}_{B,q,t} + \tilde{\boldsymbol{\beta}}_{B,q,t} - \boldsymbol{\varepsilon}_{B,q,t} \le \mathbf{A}_{B,q,t}\mathbf{P}_{B,q,t}. \quad (40)$$

Combing the constraints (33) and (39) yields:

$$\begin{aligned}&\mathbf{A}_{G,q,t}\mathbf{P}_{G,q,t} + \mathbf{A}_{B,q,t}\mathbf{P}_{B,q,t} + \mathbf{A}_{R,q,t}(\mathbf{P}_{R,q,t}-\mathbf{C}_{R,q,t}) + \mathbf{A}_{D,q,t}\mathbf{P}_{D,q,t} - \boldsymbol{\varepsilon}_{B,q,t}\\&\le \mathbf{A}_{G,q,t}\mathbf{P}_{G,q,t} + \left(\tilde{\boldsymbol{\alpha}}_{B,q,t}\widetilde{\mathbf{P}}_{B,q,t} + \tilde{\boldsymbol{\beta}}_{B,q,t}\right) + \mathbf{A}_{R,q,t}(\mathbf{P}_{R,q,t}-\mathbf{C}_{R,q,t}) + \mathbf{A}_{D,q,t}\mathbf{P}_{D,q,t}\\&\le \mathbf{P}_{F,q,t}^{\max} - \boldsymbol{\varepsilon}_{B,q,t},\end{aligned} \quad (41)$$

i.e., the constraint (5) holds.

Similarity, the existence of the constraint (6) can be checked based on the constraints (34) and (40). ∎

**Remark 3.** Denote the number of groups as $K_q$. In power industries, $K_q$ can be a small positive integer although multiple tie-lines can exist between different regional power systems. In other words, $K_q$ is usually much smaller than $n_B$. Consequently, the dimension of $\Omega_{q,t}$ decreases from $n_B+1$ to $K+1$ based on (29) if $\widetilde{\Omega}_{q,t}$ is employed to estimate $\Omega_{q,t}$. ∎

As demonstrated in Proposition 2 and Remark 3, the task of calculating $\mathbf{\Omega}_{q,t}$ comes down to the calculation of $\widetilde{\mathbf{\Omega}}_{q,t}$ that is in the domain of ($\widetilde{\mathbf{P}}_{B,q,t}$, $z_{q,t}$). This calculation can be achieved by the following vertex search method:

Step 1: Subjected to $\widetilde{\mathbf{Y}}_{q,t}$, several LP problems that separately explore the minimum and maximum of each element in ($\widetilde{\mathbf{P}}_{B,q,t}$, $z_{q,t}$) are solved. Denote optimal solutions associated with ($\widetilde{\mathbf{P}}_{B,q,t}$, $z_{q,t}$) as a set $\mathbf{V}_0$.

Step 2: Points in $\mathbf{V}_0$ represent a polytope $\mathbf{R}$. Denote the half-space representation of the polytope $\mathbf{R}$ as $\mathbf{A}_{q,t}\widetilde{\mathbf{P}}_{B,q,t}+\mathbf{B}_{q,t}z_{q,t} \leqslant \mathbf{C}_{q,t}$, where $\mathbf{A}_{q,t}$ and $\mathbf{B}_{q,t}$ are coefficient matrices and $\mathbf{C}_{q,t}$ is a coefficient vector.

Step 3: For each facet of the polytope $\mathbf{R}$, the following LP problem is solved:

$$\max_{\widetilde{\mathbf{P}}_{B,q,t}, z_{q,t}, \mathbf{x}_{q,t}, \boldsymbol{\theta}_{B,q,t}, \mathbf{P}_{B,q,t}} \mathbf{A}_{q,t}^{(k)}\widetilde{\mathbf{P}}_{B,qt} + \mathbf{B}_{q,t}^{(k)}z_{q,t}, \quad (42)$$

$$\text{s.t.} \quad (\widetilde{\mathbf{P}}_{B,q,t}, z_{q,t}, \mathbf{x}_{q,t}, \boldsymbol{\theta}_{B,q,t}, \mathbf{P}_{B,q,t}) \in \widetilde{\mathbf{Y}}_{q,t}, \quad (43)$$

where $\mathbf{A}_{q,t}^{(k)}$ and $\mathbf{B}_{q,t}^{(k)}$ are the $k^{th}$-row matrices in $\mathbf{A}_{q,t}$ and $\mathbf{B}_{q,t}$, respectively.

The intuitive explanation of the LP problem (42)-(43) is to move the facet $\mathbf{A}_{q,t}^{(k)}\widetilde{\mathbf{P}}_{B,q,t}+\mathbf{B}_{q,t}^{(k)}z_{q,t}$ as far as possible away from the center of $\widetilde{\mathbf{\Omega}}_{q,t}$. Once the LP problem (42)-(43) is solved, denote all optimal solutions associated with ($\widetilde{\mathbf{P}}_{B,q,t}$, $z_{q,t}$) as a set $\mathbf{V}_{new}$. Let $\{\mathbf{V}_{new} \cup \mathbf{V}_0\} \rightarrow \mathbf{V}_0$. This constructs a new polytope $\mathbf{R}_{new}$.

Step 4: The difference between $\mathbf{R}$ and $\mathbf{R}_{new}$ can be measured by shape changes. To quantify shape changes, the difference between the volume of $\mathbf{R}$ and the volume of $\mathbf{R}_{new}$ is calculated [15]. Once the difference is smaller than a given threshold, the feasible region $\widetilde{\mathbf{\Omega}}_{q,t}$ can be regarded as the polytope $\mathbf{R}_{new}$; otherwise, let $\mathbf{R}_{new} \rightarrow \mathbf{R}$ and go back to Step 3.

Denote the final half-space representation of $\widetilde{\mathbf{\Omega}}_{q,t}$ as

$$\widetilde{\mathbf{\Omega}}_{q,t} \triangleq \left\{ (\widetilde{\mathbf{P}}_{B,q,t}, z_{q,t}) \middle| \mathbf{A}_{B,q,t}^{(P)} \widetilde{\mathbf{P}}_{B,t} + \mathbf{A}_{B,q,t}^{(z)} z_{q,t} \leq \mathbf{B}_{q,t}^{(Pz)} \right\}, \quad (44)$$

where $\mathbf{A}_{B,q,t}^{(P)}$ and $\mathbf{A}_{B,q,t}^{(z)}$ are constant matrices obtained in the four steps mentioned above; $\mathbf{B}_{q,t}^{(Pz)}$ is a constant vector obtained in the four steps mentioned above.

**Remark 4.** The calculation of $\widetilde{\mathbf{\Omega}}_{q,t}$ is a projection of $\widetilde{\mathbf{Y}}_{q,t}$ from the domain of ($\widetilde{\mathbf{P}}_{B,q,t}$, $z_{q,t}$, $\mathbf{x}_{q,t}$, $\boldsymbol{\theta}_{B,q,t}$) to the domain of ($\widetilde{\mathbf{P}}_{B,q,t}$, $z_{q,t}$). The four steps mentioned above pave an way to explore $\widetilde{\mathbf{\Omega}}_{q,t}$ by finding its vertices. Particularly, $z_{q,t}$ at vertices are either $z_{q,t}= \mathbf{e}_{R,q,t}^T \mathbf{C}_{R,q,t}$ or $z_{q,t}= \mathbf{e}_{R,q,t}^T \mathbf{P}_{R,q,t}$ based on (11) that is incorporated in $\widetilde{\mathbf{Y}}_{q,t}$. Also, each vertex is found in the Step 3 by solving the LP problem (42)-(43) that moves a hyperplane as far as possible away from the center of $\widetilde{\mathbf{\Omega}}_{q,t}$. Consequently, $z_{q,t}$ at certain vertices reaches the infimum of $\mathbf{e}_{R,q,t}^T \mathbf{C}_{R,q,t}$. This preserves the information of the objective function (14) in our tie-line security region. ∎

Once the tie-line security region is obtained based on the four steps mentioned above, the coordination in an interconnected power system can be achieved under a decentralized and non-iterative framework, as will be elaborated in Sec. V.

## V. RENEWABLE ACCOMMODATIONS VIA TIE-LINE SECURITY REGIONS

In this section, the tie-line security region obtained in Sec. IV will be utilized for renewable accommodations in an interconnected power system under a decentralized and non-iterative framework. To achieve this goal, the coupling constraints (16) will be explicitly formulated over the tie-line security region in Sec. V-A. Furthermore, an LP problem to reduce renewable curtailments by coordinating tie-line power and border angles will be established within the tie-line security regions of different regional power systems.

### A. Reformulation of the coupling constraints (16) over the tie-line security region

For the regional power system $q$, its $\mathbf{P}_{B,q,t}$ and $\boldsymbol{\theta}_{B,q,t}$ are incorporated into the coupling constraints (16). The term $\mathbf{P}_{B,q,t}$ has been involved in the tie-line security region $\widetilde{\mathbf{\Omega}}_{q,t}$ by $\widetilde{\mathbf{P}}_{B,q,t}$. However, the term $\boldsymbol{\theta}_{B,p,t}$ is not involved in the tie-line security region $\widetilde{\mathbf{\Omega}}_{q,t}$. Once $\boldsymbol{\theta}_{B,q,t}$ is formulated as a function of $\widetilde{\mathbf{P}}_{B,q,t}$, the coupling constraints can be easily expressed over the tie-line security region. Proposition 3 is presented to construct the relationship between $\boldsymbol{\theta}_{B,q,t}$ and $\widetilde{\mathbf{P}}_{B,q,t}$.

**Proposition 3.** Denote vertices of $\widetilde{\mathbf{\Omega}}_{q,t}$ as the set $\mathbf{V}$ that is described as follows:

$$\mathbf{V} \triangleq \left\{ (\widetilde{\mathbf{P}}_{B,q,t}^{(i)}, z_{q,t}^{(i)}), i = \{1,2,...,S_{q,t}\} \right\}, \quad (45)$$

where $S_{q,t}$ is the number of vertices; $(\widetilde{\mathbf{P}}_{B,q,t}^{(i)}, z_{q,t}^{(i)})$ is the $i^{th}$ vertex.

All vertices in the set $\mathbf{V}$ are found by repeatedly solving the LP problem (42)-(43). For the $i^{th}$ vertex obtained by solving the LP problem (42)-(43), a corresponding point ($\mathbf{x}_{q,t}^{(i)}$, $\boldsymbol{\theta}_{B,q,t}^{(i)}$, $\mathbf{P}_{B,q,t}^{(i)}$) exists within $\widetilde{\mathbf{Y}}_{q,t}$.

Based on the definitions mentioned above, the relationship between $\boldsymbol{\theta}_{B,q,t}$ and $\widetilde{\mathbf{P}}_{B,q,t}^{(i)}$ can be expressed in the following.

$$\widetilde{\mathbf{P}}_{B,q,t} = \sum_{i=1}^{S_{q,t}} \lambda_{q,t,i} \widetilde{\mathbf{P}}_{B,q,t}^{(i)}, \quad (46)$$

$$\boldsymbol{\theta}_{B,q,t} = \sum_{i=1}^{S_{q,t}} \lambda_{q,t,i} \boldsymbol{\theta}_{B,q,t}^{(i)}, \quad (47)$$

$$z_{q,t} = \sum_{i=1}^{S_{q,t}} \lambda_{q,t,i} z_{q,t}^{(i)}, \forall q, \forall t, \quad (48)$$

$$\lambda_{q,t,i} \geq 0, i = \{1,2,...,S_{q,t}\}, \sum_{i=1}^{S_{q,t}} \lambda_{q,t,i} = 1. \quad (49)$$

∎

**Proof.** This proof is completed by checking the existence of $\widetilde{\mathbf{Y}}_{q,t}$ when the relationship in (46)-(49) is given. Considering the convexity of $\widetilde{\mathbf{Y}}_{q,t}$, each point ($\widetilde{\mathbf{P}}_{B,q,t}$, $z_{q,t}$, $\mathbf{x}_{q,t}$, $\boldsymbol{\theta}_{B,q,t}$, $\mathbf{P}_{B,q,t}$) within $\widetilde{\mathbf{Y}}_{q,t}$ can be a convex combination of vertices, i.e.,

$$\left[ (\widetilde{\mathbf{P}}_{B,q,t})^T \quad z_{q,t} \quad (\mathbf{x}_{q,t})^T \quad (\boldsymbol{\theta}_{B,q,t})^T \quad (\mathbf{P}_{B,q,t})^T \right]^T$$
$$= \sum_{i=1}^{S_{q,t}} \lambda_{q,t,i} \left[ (\widetilde{\mathbf{P}}_{B,q,t}^{(i)})^T \quad z_{q,t}^{(i)} \quad (\mathbf{x}_{q,t}^{(i)})^T \quad (\boldsymbol{\theta}_{B,q,t}^{(i)})^T \quad (\mathbf{P}_{B,q,t}^{(i)})^T \right]^T, \quad (50)$$

where $\sum_{i=1}^{S_{q,t}} \lambda_{q,t,i} = 1$ and $\lambda_{q,t,i} \geq 0$ for $i=\{1,2,…,S_{q,t}\}$.

The relationship in (46)-(49) holds due to (50). ∎

Based on Proposition 3, the coupling constraint between different regional power systems can be formulated as the combinations of (16) and (46)-(49). Such a reformulation is over the tie-line security region.

### B. Renewable accommodations via tie-line security region

The following LP problem is designed to reduce renewable curtailments via tie-line security regions:

$$\min_{\{z_{q,t},\tilde{\mathbf{P}}_{B,q,t},\mathbf{P}_{B,q,t},\boldsymbol{\theta}_{B,q,t},\lambda_{q,t,i}\},\forall q,\forall t,\forall i} \sum_{q=1}^{n_Q}\sum_{t=1}^{n_T} z_{q,t}, \quad (51)$$

s.t. $\mathbf{P}_{B,q,t} = \mathbf{P}_{B,p,t} = \mathbf{D}_{B,p}\boldsymbol{\theta}_{B,p,t} + \mathbf{D}_{B,q}\boldsymbol{\theta}_{B,q,t}, \forall q,\forall p,\forall t, p\neq q,$ (52)

$$z_{q,t} = \sum_{i=1}^{S_{q,t}} \lambda_{q,t,i} z_{q,t}^{(i)}, \forall q, \forall t, \quad (53)$$

$$\tilde{\mathbf{P}}_{B,q,t} = \sum_{i=1}^{S_{q,t}} \lambda_{q,t,i} \tilde{\mathbf{P}}_{B,q,t}^{(i)}, \forall q, \forall t, \quad (54)$$

$$\boldsymbol{\theta}_{B,q,t} = \sum_{i=1}^{S_{q,t}} \lambda_{q,t,i} \boldsymbol{\theta}_{B,q,t}^{(i)}, \forall q, \forall t, \quad (55)$$

$$\lambda_{q,t,i} \geq 0, i = \{1,2,...,S_{q,t}\}, \forall q, \forall t, \quad (56)$$

$$\sum_{i=1}^{S_{q,t}} \lambda_{q,t,i} = 1, \forall q, \forall t, \quad (57)$$

$$\tilde{\mathbf{P}}_{B,q,t} = \tilde{\mathbf{A}}_{B,q,t} \mathbf{P}_{B,q,t}, \forall q, \forall t, \quad (58)$$

$$\mathbf{P}_{B,q,t}^{\min,re} \leq \mathbf{P}_{B,q,t} \leq \mathbf{P}_{B,q,t}^{\max,re}. \quad (59)$$

The objective function is to reduce renewable curtailments because $z_{q,t}$ is the variable associated with the total renewable curtailments in each regional power system, as defined in Sec. III. The constraints (52)-(57) over the tie-line security region are the coupling constraints between different regional power systems, as mentioned in Sec. V-A. The constraint (58) is the mapping between $\tilde{\mathbf{P}}_{B,q,t}$ and $\mathbf{P}_{B,q,t}$ when $\mathbf{P}_{B,q,t}$ is restricted by its upper and lower limits (59), as mentioned in Sec. IV.

By solving the LP problem (51)-(59), $\mathbf{P}_{B,q,t}$ and $\boldsymbol{\theta}_{B,q,t}$ can be fixed as $(\mathbf{P}_{B,q,t}^*, \boldsymbol{\theta}_{B,q,t}^*)$. Given $(\mathbf{P}_{B,q,t}^*, \boldsymbol{\theta}_{B,q,t}^*)$, the following LP problem can be solved by the regional power system $q$ for its decisions:

$$\min_{\{\mathbf{x}_{q,t}, z_{q,t}\}, \forall t} \sum_{t=1}^{n_T} z_{q,t}, \quad (60)$$

s.t. $(\mathbf{x}_{q,t}, \mathbf{P}_{B,q,t}^*, \boldsymbol{\theta}_{B,q,t}^*, z_{q,t}) \in \mathbf{X}_{q,t}, \forall t. \quad (61)$

**Remark 5**. The LP problem (51)-(59) is solved over the tie-line security region. Consequently, the coordination to reduce renewable curtailments among different regional power systems is achieved under a decentralized and non-iterative framework. ∎

**Remark 6**. The decisions made by solving the LP problem (51)-(59) guarantee the feasibility of the original constraints (3)-(13). This arises from that the constraints (51)-(59) are designed from Propositions 1-3 whose preconditions are the feasibility of the constraints (3)-(13). ∎

The flowchart of renewable accommodations based on the tie-line security region is shown in Fig. 1.

## VI. CASE STUDIES

The presented methods are verified in the IEEE 9-bus system, a 661-bus utility system, and a five-region system. Firstly, the presented fast calculation method is verified in the IEEE 9-bus test system in Sec. V-A. Secondly, the computational performance of the presented fast calculation method is compared with the representatives in a 661-bus utility system in Sec. V-B. Finally, the presented method for renewable accommodations via the tie-line security region is verified in a five-region system in Sec. V-C.

All numerical results are calculated with MATLAB 2015b and performed on a desktop computer with Intel (R) Core (TM) i5-4460 CPU @ 3.20GHz 8.00G RAM. All LP problems are solved via YALMIP and CPLEX.

### A. Verification of the presented fast calculation method

The fast calculation method presented in Sec. IV relies on the decomposition of time periods and aggregation of tie-line power. They are verified in the IEEE 9-bus test system in this subsection.

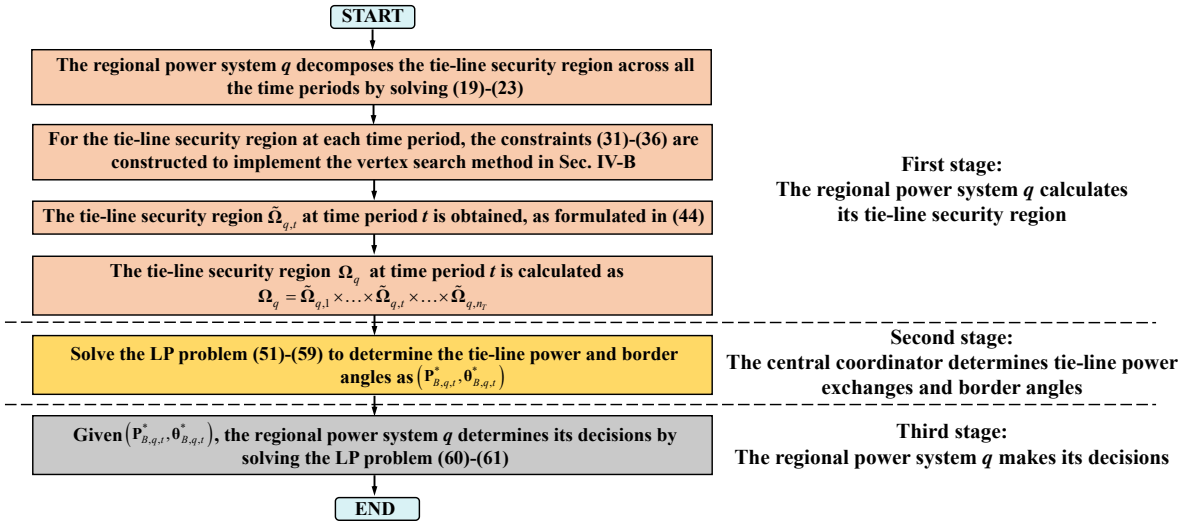

Fig. 1 Flowchart of the presented methods

➢ *Verification of the decomposition of time periods*

The decomposition of time periods presented in Sec. IV-A is verified in the IEEE 9-bus test system with two time periods [21]. Each border bus is connected with a tie-line.

As demonstrated in Sec. IV.A, the complete tie-line security region across time periods 1 and 2 is a Cartesian product of the tie-line security region at each time period. To verify the legality of the decomposition, 10000 points are randomly selected from our feasible region across two time periods. For each point, the feasibility of the original constraints (3)-(14) is checked by solving the OP1 problem. Numerical experiments are listed in Table 1. It can be found that time coupling is not violated in the presence of our decomposing time periods

Table 1 Numerical results to verify the decomposition

|  | Feasible point | Infeasible point |
| --- | --- | --- |
| Number | 10000 | 0 |

➢ *Verification of aggregating tie-line power*

The aggregation of tie-line power presented in Sec IV-B is verified in the IEEE 9-bus test system under a single time period [21]. Four border buses 1, 3, 7, and 9 are respectively connected with four tie-lines. Four tie-lines are categorized into two groups. Tie-lines connected to buses 1 and 9 are in one group, while tie-lines connected to buses 3 and 7 are in another group. Furthermore, the tie-line security region is calculated, as shown in Fig. 2.

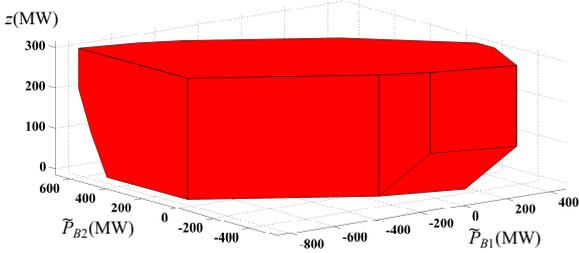

Fig. 2 Tie-line security region after aggregation. $\widetilde{P}_{B1}$ is the aggregation of tie-lines at buses 3 and 7. $\widetilde{P}_{B2}$ is the aggregation of tie-lines at buses 1 and 9. $z$ is the additional variable associated with renewable curtailments.

To verify the tie-line security region shown in Fig. 2, 10000 points are randomly selected. For each selected point, the feasibility of the original constraints (3)-(14) is checked by solving the OP1 problem. The numbers of feasible points and infeasible points are listed in Table 2. As shown in Table 2, all selected points are feasible. In other words, the aggregation of tie-line power presented in Sec. IV-B can guarantee the feasibility of the original constraints (3)-(14).

Table 2 Numerical results to verify the aggregation

|  | Feasible point | Infeasible point |
| --- | --- | --- |
| Number | 10000 | 0 |

B. *Comparison of computational performance*

In this subsection, the computational performance of the following five methods is compared:

**M1**: The fast calculation method presented in Sec. IV.
**M2**: The method based on multi-parametric programming in [12].
**M3**: The method based on vertex search in [14].
**M4**: The method based on the simple and heuristic decomposition in [15].
**M5**: The method based on Fourier-Motzkin elimination [16].

The five methods mentioned above are tested in a 661-bus utility system with six time periods. Three border buses are correspondingly connected by three tie-lines. The computational time of different methods is listed in Table 3.

Table 3 Computational time of five methods

| Method | M1 | M2 | M3 | M4 | M5 |
| --- | --- | --- | --- | --- | --- |
| Time (seconds) | 11.3 | >3600 | >3600 | 2908.6 | >3600 |

In this test system, the M1 method (i.e., our methods) has the least time because the computational burden brought by multiple time periods and border buses are resolved. The M2, M3 and M5 method cannot obtain the tie-line security region within one hour because of the heavy computational burden, as analyzed in Sec. II. Particularly, the M4 method obtains its result at the expense of 2908.6 seconds. This arises from that the computational burden the computational burden brought by multiple time periods is handled by simply decomposing as the union of a few lower-dimension polytopes at certain combinations of different time periods, while the computational burden brought by multiple border buses still remains.

Furthermore, the feasibility of the tie-line security regions obtained by the M1 and M4 methods are compared. The same way in Table 2 is employed to check their feasibility. The numerical results are listed in Table 4. As shown in Table 4, the M1 method can guarantee the feasibility, as have been shown in Sec. VI-A. As for the M3 method, 2342 points selected from its tie-line security region are infeasible. This arises from its simple and heuristic decomposition of time periods, as also demonstrated in Fig. 5 in [15].

Table 4 Numerical results of checking the feasibility

| Method | Number of total selected points | Number of feasible points | Number of infeasible points |
| --- | --- | --- | --- |
| M0 | 10000 | 10000 | 0 |
| M3 | 10000 | 7658 | 2342 |

C. *Verification of renewable accommodations via the tie-line security region*

In this subsection, a five-region system that is constructed from a 661-bus utility system with 24 time periods is employed to verify renewable accommodations via the tie-line security region. The diagram of such a five-region system is shown in Fig. 3.

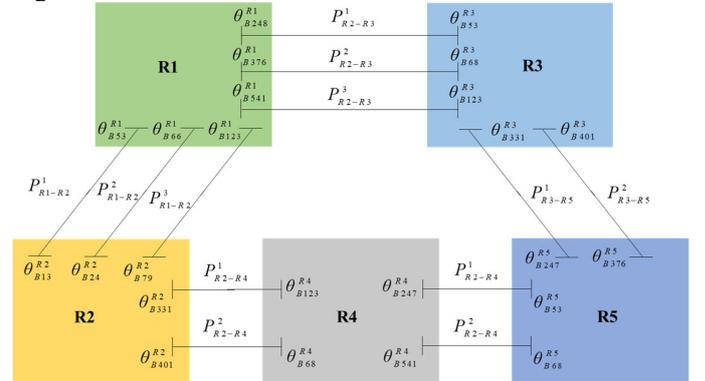

Fig. 3 Diagram of a five-region system

In the five-region system, renewable accommodations under the following two different frameworks are compared:

**F1**: No power exchanges among different regional power systems.

**F2**: The decentralized framework and non-iterative framework based on Fig. 1, i.e., renewable accommodations are implemented via our tie-line security region.

Two frameworks are implemented under 50 renewable scenarios. Their renewable accommodations are compared in Fig. 4. It can be found that renewable accommodations under F2 are higher than those under F1. This revisits the necessity of accommodating renewables via tie-line power exchanges.

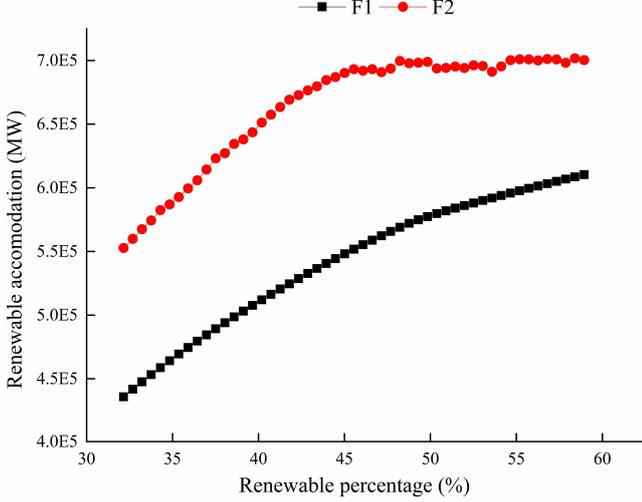

Fig. 4 Renewable curtailments under 50 renewable scenarios

Furthermore, the computational time under F2 is listed in Fig. 5. Its major computational burden lies in two parts: one is to calculate tie-line security regions of R1-R5, and the other one is the coordination by solving (51)-(59) and (60)-(61). As shown in Fig. 5, the computational time of calculating a tie-line security region is not more than 700 seconds under the tested 50 renewable scenarios. As for the coordination by solving (51)-(59) and (60)-(61), its computational time is not more than three seconds. Consequently, it is believed that our method can provide an opportunity to coordinate renewable accommodations via tie-line security regions when a limited time is required in power industries.

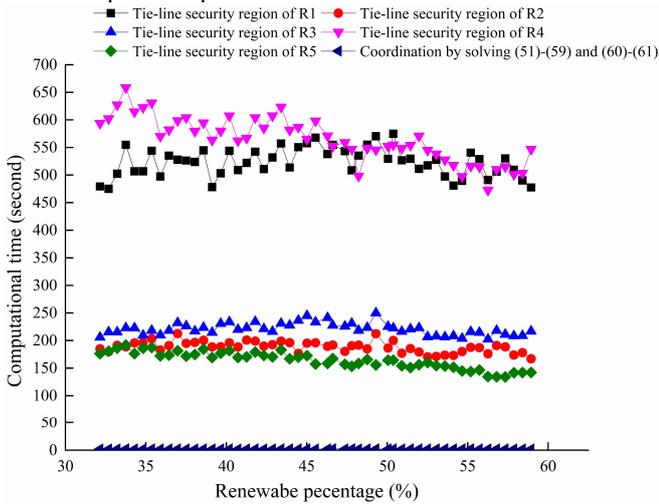

Fig. 5 Computational time under F2

## VII. CONCLUSIONS

A tie-line security region is essentially a high-dimension polytope when multiple time periods and multiple border buses inherently in power system operations are involved. Its calculation suffers a considerable computational burden in the existing methods. Consequently, the fast calculation of a tie-line security region with multiple time periods and multiple border buses is studied in this paper, together with its application to facilitate renewable accommodation in an interconnected power system.

The computational burden brought by multiple time periods is avoided by leverages dispatch levels with which ramp rate constraints always hold even if ramp rate constraints are neglected. Consequently, the tie-line security region across all the time periods can be decomposed as a Cartesian production of the lower-dimension tie-line security region at each time period.

Furthermore, for the tie-line security region at each time period, the computational burden brought by multiple border buses is alleviated by aggregating tie-line power. The explicit mapping function and constraints are designed to guarantee the feasibility of the original tie-line power recovered from the aggregated tie-line power.

Also, an additional dimension associated with minimum renewable curtailments is incorporated in the tie-line security region. This facilitates the coordination of renewable accommodations via the tie-line security region. Based on the tie-line security region, a linear programming problem is designed for renewable accommodations in an interconnected power system under a decentralized and non-iterative framework. Case studies based on the IEEE 9-bus system, a 661-bus utility system and a five-region system corroborate the effectiveness of the presented methods.

## APPENDIX

The $l^{th}$ element in $\mathbf{A}_{B,q,t}\mathbf{P}_{B,q,t}$ in (5)-(6) can be separated based on the groups of tie-line power, as described below.

$$\mathbf{A}^l_{B,q,t}\mathbf{P}_{B,q,t} = \sum_{j=1}^{K}\left(\sum_{i=1}^{n_j} \alpha^{l,j_i}_{B,q,t} P^{j_i}_{B,q,t}\right), \quad (A1)$$

where $\mathbf{A}^l_{B,q,t}$ is the $l^{th}$-row matrix of $\mathbf{A}_{B,q,t}$; $\alpha^{l,j_i}_{B,q,t}$ is the coefficient in $\mathbf{A}^l_{B,q,t}$ and is associated with $P^{j_i}_{B,q,t}$.

In this paper, the term $\sum_{i=1}^{n_j} \alpha^{l,j_i}_{B,q,t} P^{j_i}_{B,q,t}$ is approximated by $\tilde{\alpha}^{l,j}_{B,q,t} P^{j}_{B,q,t} + \tilde{\beta}^{l,j}_{B,q,t}$, where $P^{j}_{B,q,t}$ is the aggregation of $\sum_{i=1}^{n_j} P^{j_i}_{B,q,t}$ as shown in (29). In other words, $\mathbf{A}_{B,q,t}\mathbf{P}_{B,q,t}$ in (5)-(6) is approximated by $\tilde{\boldsymbol{\alpha}}_{B,q,t}\widetilde{\mathbf{P}}_{B,q,t}+\tilde{\boldsymbol{\beta}}_{B,q,t}$ by calculating $\tilde{\alpha}^{l,j}_{B,q,t}$ and $\tilde{\beta}^{l,j}_{B,q,t}$. Their error can be bounded by the error bound $\boldsymbol{\varepsilon}_{B,q,t}$ whose dimension is the number of rows in $\mathbf{A}_{B,q,t}$. The $l^{th}$ element in $\boldsymbol{\varepsilon}_{B,q,t}$ is denoted as $\varepsilon^l_{B,q,t}$, which can be calculated in the following:

$$\varepsilon^l_{B,q,t} = \sum_{j=1}^{K}\varepsilon^{l,j}_{B,q,t}, \quad (A2)$$

where $\varepsilon^{l,j}_{B,q,t}$ is calculated in the following:

$$\varepsilon_{B,q,t}^{l,j} = \min_{\tilde{\alpha}_{B,q,t}^{l,j}, \tilde{\beta}_{B,q,t}^{l,j}} \max_{P_{B,q,t}^{(j_i)}} \left| \sum_{i=1}^{n_j} \left( \alpha_{B,q,t}^{l,j_i} P_{B,q,t}^{(j_i)} \right) - \left( \tilde{\alpha}_{B,q,t}^{l,j} \sum_{i=1}^{n_j} \left( P_{B,q,t}^{(j_i)} \right) + \tilde{\beta}_{B,q,t}^{l,j} \right) \right|, \quad (A3)$$

s.t. Constraints (35). (A4)

For convenience of discussion, (A3)-(A4) are equivalently formulated as the following problem:

$$\varepsilon_0 = \min_{a_0, b_0} \max_{x_s, \forall s} \left| \sum_{s=1}^{S} a_s x_s - \left( a_0 \sum_{s=1}^{S} x_s + b_0 \right) \right|, \quad (A5)$$

s.t. $0 \leq x_s \leq x_s^{\max}, \forall s.$ (A6)

Note that the lower bound $x_s$ is zero. This can be achieved by simple linear transformation. In addition, the coefficients in (A5) are ranked by the following rule:

$$a_1 \leq a_2 \leq \ldots \leq a_n, \quad (A7)$$

The global optimal solution of (A5)-(A6) can be explicitly formulated as follows (see Theorem 2 in [20]):

$$a_0^* = a_c, \quad (A8)$$

$$b_0^* = \frac{1}{2} \sum_{s=1}^{S} \left( a_s - a_0^* \right) x_s^{\max}, \quad (A9)$$

$$\varepsilon_0^* = \frac{1}{2} \left[ \sum_{s=c}^{n} a_s x_s^{\max} - \sum_{s=1}^{c-1} a_s x_s^{\max} + a_0^* \left( \sum_{s=1}^{c-1} x_s^{\max} - \sum_{s=c}^{n} x_s^{\max} \right) \right], \quad (A10)$$

where the index $c$ is determined in the following way:

$$\lambda_c \leq 0, \lambda_{c+1} \geq 0, \quad (A11)$$

where $\lambda_j$ in (A11) is denoted as

$$\lambda_j = \sum_{s=1}^{j-1} x_s^{\max} - \sum_{s=j}^{n} x_s^{\max}. \quad (A12)$$

Based on (A2)-(A3) and (A8)-(A10), $\tilde{\alpha}_{B,q,t}^{l,j}$ and $\tilde{\beta}_{B,q,t}^{l,j}$ can be easily obtained to construct the constant $\tilde{\boldsymbol{\alpha}}_{B,q,t}$, $\tilde{\boldsymbol{\beta}}_{B,q,t}$ and $\boldsymbol{\varepsilon}_{B,q,t}$.